\documentclass[amssymb,aps,floatfix,floats,preprintnumbers,twocolumn,superscriptaddress,nofootinbib]{revtex4}
\usepackage{rcs}
\usepackage{color,graphicx}
\usepackage{dcolumn}
\usepackage{bm}
\usepackage[normalem]{ulem}
\usepackage{chngcntr}
\usepackage[thinspace,thinqspace]{SIunits}
\usepackage{amsmath}
\usepackage{natbib}
\usepackage{doi}
\usepackage{hyperref}

\newcommand{\muB}{$\mu_{\textrm{B}}$}
\newcommand{\muo}{$\mu_{\textrm{0}}$}

\newcommand{\CRA}{CeRh$_{2}$As$_{2}$}

\newcommand{\Tc}{$T_{\textrm{c}}$}
\newcommand{\To}{$T_{\textrm{0}}$}

\newcommand{\Ho}{$H_{\textrm{0}}$}
\newcommand{\Hstar}{$H^{\textrm{*}}$}

\begin{document}
\preprint{APS/123-QED}
\title{The phase diagram of \CRA\ for out-of-plane magnetic field}

\author{P. Khanenko}
\email[Corresponding author:~]{pavlo.khanenko@cpfs.mpg.de}
\affiliation{Max Planck Institute for Chemical Physics of Solids, D-01187 Dresden, Germany}
\author{J. F. Landaeta}
\author{S. Ruet}
\author{T. L\"uhmann}
\affiliation{Max Planck Institute for Chemical Physics of Solids, D-01187 Dresden, Germany}
\author{K. Semeniuk}
\affiliation{Max Planck Institute for Chemical Physics of Solids, D-01187 Dresden, Germany}
\affiliation{Technical University Dresden, Institute for Solid State and Materials Physics, 01062 Dresden, Germany}
\author{M. Pelly}
\author{A. W. Rost}
\affiliation{Scottish Universities Physics Alliance, School of Physics and Astronomy, University of St. Andrews,
St. Andrews, Fife KY16 9SS, United Kingdom}
\author{G. Chajewski}
\author{D. Kaczorowski}
\affiliation{Institute of Low Temperature and Structure Research, Polish Academy of Sciences, Ok\'olna 2, 50-422 Wroc{\l}aw, Poland}
\author{C. Geibel}
\affiliation{Max Planck Institute for Chemical Physics of Solids, D-01187 Dresden, Germany}
\author{S. Khim}
\affiliation{Max Planck Institute for Chemical Physics of Solids, D-01187 Dresden, Germany}
\author{E. Hassinger}
\affiliation{Technical University Dresden, Institute for Solid State and Materials Physics, 01062 Dresden, Germany}
\author{M. Brando}
\email[Corresponding author:~]{manuel.brando@cpfs.mpg.de}
\affiliation{Max Planck Institute for Chemical Physics of Solids, D-01187 Dresden, Germany}
\date{\today}
\begin{abstract}
The heavy-fermion superconductor \CRA\ (\Tc\ = 0.35\,K) shows two superconducting (SC) phases, SC1 and SC2, when a magnetic field is applied parallel to the $c$ axis of the tetragonal unit cell. All experiments to date indicate that the change in SC order parameter detected at \muo$H^{*} \approx 4$\,T is due to strong Rashba spin-orbit coupling at the Ce sites caused by the locally non-centrosymmetric environments of the otherwise globally centrosymmetric crystalline structure. Another phase (phase I) exists in this material below \To\ = 0.54\,K. In a previous specific heat study [K. Semeniuk \textit{et al.} Phys. Rev. B, \textbf{107}, L220504 (2023)] we have shown that phase I persists up to a field \muo$H_{0} \approx 6$\,T, larger than $H^{*}$. From thermodynamic arguments we expected the phase-I boundary line to cross phase SC2 at a tetracritical point. However, we could not find any signature of the phase-I line inside the SC2 phase and speculated that this was due to the fact that the $T_{0}(H)$ line is almost perpendicular to the $H$ axis and, therefore, invisible to $T$-dependent measurements. This would imply a weak competition between the two order parameters. Here, we report magnetic field dependent measurements of the magnetostriction and ac-susceptibility on high-quality single crystals. We see clear evidence of the singularity at $H_{0}$ inside the SC2 phase and confirm our previous prediction. Furthermore, we observe the transition across the $T^{*}(H)$ line in $T$-dependent specific heat measurements, which show that the $T^{*}(H)$ line is not perpendicular to the field axis, but has a positive slope. Our work supports recent $\mu$SR results which suggest coexistence of phase I with superconductivity.
\end{abstract} 
\maketitle
\section{Introduction}
Multi-phase superconductivity is very rare. It is realized in superfluid $^{3}$He~\cite{leggett1975} and only in a few U-based unconventional superconductors, i.e., UPt$_{3}$~\cite{fisher1989,bruls1990,adenwalla1990}, thorium-doped UBe$_{13}$~\cite{ott1985} and UTe$_{2}$~\cite{ran2019,braithwaite2019,aoki2020}. Despite many years of experimental and theoretical efforts there is no model that can fully describe the behavior of these superconductors~\cite{taillefer2010}. 

The recently discovered heavy-fermion (HF) superconductor \CRA\ is the only established non-U-based multi-phase unconventional superconductor to date~\cite{khim2021}. Importantly, its behavior can be well described by theories based on strong Rashba spin-orbit coupling at the Ce sites caused by the locally non-centrosymmetric environments of the otherwise globally centrosymmetric crystalline structure~\cite{yoshida2012,sigrist2014}. Thus, studying \CRA\ opens up a unique opportunity for better understanding unconventional superconductivity, in particular in locally non-centrosymmetric systems. 

\begin{figure}[t]
	\begin{center}
		\includegraphics[width=\columnwidth]{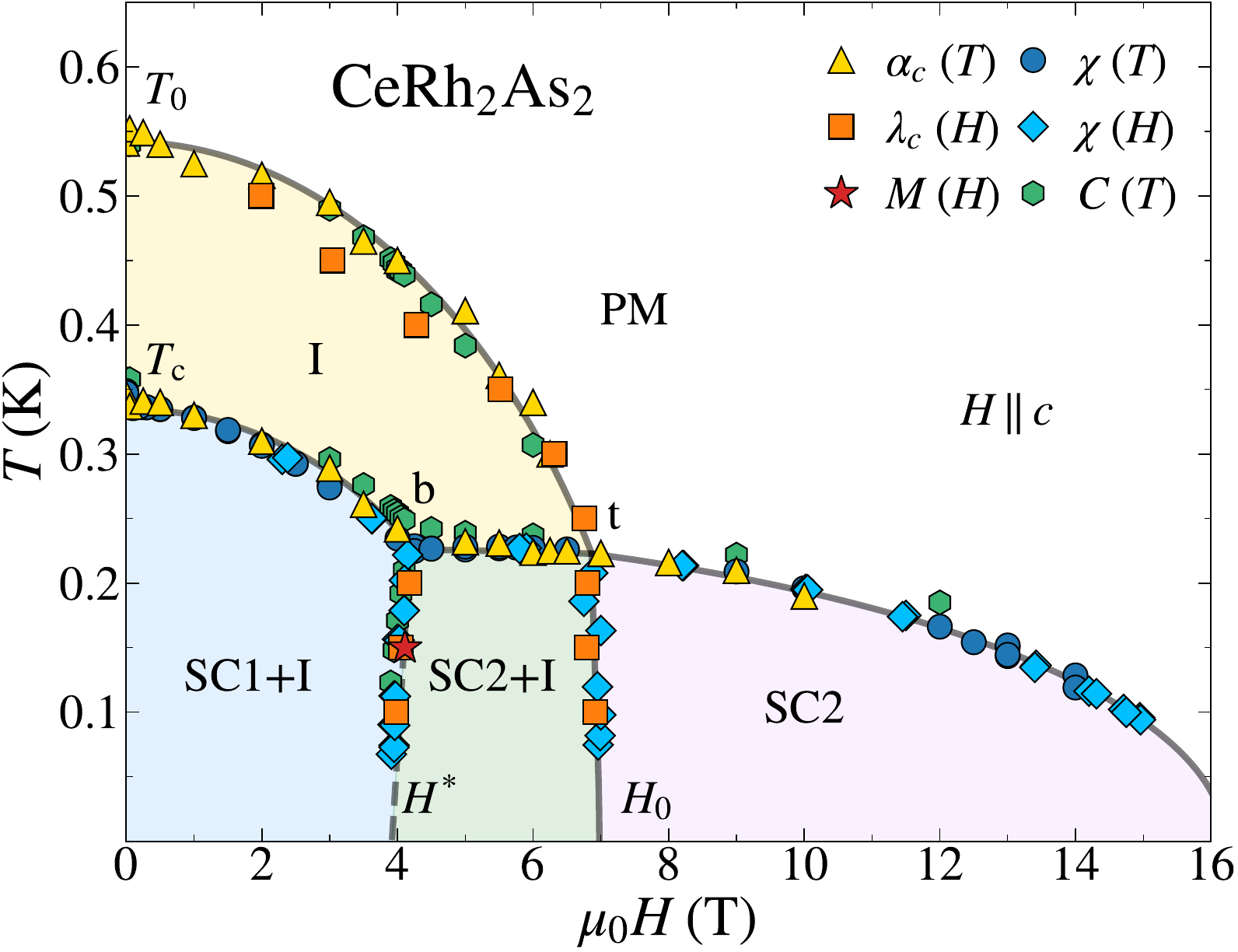}
		\caption{Refined $H-T$ phase diagram of \CRA\ for $H \parallel c$-axis. Points are taken from specific heat $C(T)$, magnetic susceptibility $\chi(T,H)$, linear thermal expansion coefficient $\alpha_{c}(T)$, linear magnetostriction coefficient $\lambda_{c}(H)$ and magnetization $M(H)$ experiments.  The letter 'b' and 't' indicate the 'bicritical' and 'tetracritical' points discussed in the text. The solid gray lines fit to the GL model described in the text. The dashed gray line is a guide for the eye. Since $\mu$SR experiments suggest coexistence of phase I with superconductivity, and the slight change of slope of the $T_{0}(H)$ line at 't' implies weak coupling between phase I and SC2, we have labeled the mixed phases, SC1+I and SC2+I.}
		\label{fig1}
	\end{center}
\end{figure}

The most striking information that connects the SC states in \CRA\ with the theories mentioned above is the shape of the magnetic $H - T$ phase diagram when a magnetic field is applied parallel to the $c$ axis of the tetragonal unit cell. A refined version of the phase diagram is shown in Fig.~\ref{fig1} and is the main result of our work presented here.  The two superconducting (SC) phases, SC1 and SC2, are separated by a weak first-order line at \muo$H^{*} \approx 4$\,T. This line was observed in field-dependent magnetization, magnetostriction and ac-susceptibility measurements~\cite{khim2021,landaeta2022a}. In contrast, for in-plane magnetic field there is only one SC phase, SC1, which is Pauli limited with a critical field of about 2\,T~\cite{khim2021,landaeta2022a,khanenko2025}. The pronounced anisotropy of the SC phases was also predicted by the same  theories~\cite{yoshida2012,sigrist2014,moeckli2018,schertenleib2021,skurativska2021,nogaki2021,ptok2021,moeckli2021,cavanagh2022,nogaki2022,ishizuka2024,nally2024,lee2024}.

As in the case of the multi-phase HF UPt$_{3}$~\cite{bruls1990}, in \CRA\ there exists another phase, labeled I. This phase is present at temperatures above superconductivity with a transition temperature \To\ $\approx 0.5$\,K~\cite{hafner2022,khanenko2025} (cf. Fig.~\ref{fig1}). Previous studies suggested that the nature of the ordering is a non-magnetic quadrupolar-density-wave (QDW) instability supported by the presence of a quasi-quartet ground state of the crystal electric field~\cite{hafner2022,christovam2023,takimoto2008}. However, NQR/NMR~\cite{kibune2022,ogata2023} as well as recent $\mu$SR~\cite{khim2025} experiments have clearly detected antiferromagnetic (AFM) order below \To. Also, the presence of AFM quasi-2d fluctuations in inelastic neutron scattering (INS) experiments~\cite{chen2024} and the analysis of the specific heat~\cite{chajewski2024} support the presence of AFM order at \To. A possible concomitant ordering of dipolar and quadrupolar degrees of freedom at \To\ is currently under debate and is strongly suggested by the in-plane field phase diagram which shows a substantial increase of \To\ with the field~\cite{schmidt2024,khanenko2025}.

One of the central questions in such systems is the nature of the interplay between the superconducting state and other ordered state, in the present case phase I. With the present results, we are able to provide the answer to one essential point, namely the sign and the strength of the interaction between the SC2 state and phase I. In case of a strong coupling between the order parameters, the symmetry of phase I might reduce the allowed symmetry possibilities for the superconducting order parameters. In a previous work, Ref.~\cite{semeniuk2023}, we have shown that the signatures of the phase I boundary line $T_{0}(H)$ disappear just before entering the SC2 phase. The highest field, at which phase I was clearly detected, was \muo$H_{0} \approx 5$\,T, i.e., a field larger than $H^{*}$. From thermodynamic arguments we predicted that $T_{0}(H)$ should enter into the SC2 phase. However, we could not find any signature of this behavior in our previous $T$-dependent measurements. We speculated that this was due to the fact that the $T_{0}(H)$ line is almost perpendicular to the $H$ axis and, therefore, invisible to $T$-dependent measurements. 

Here, by means of $H$-dependent magnetostriction and ac-susceptibility measurements on high-quality single crystals, we clarify this point and confirm our prediction. The behavior of the phase boundary lines shown in Fig.~\ref{fig1} implies that there is a weak competition between the phase-I and the SC2 order parameters. Furthermore, we observe the transition across the $T^{*}(H)$ line in $T$-dependent specific heat measurements, which show that this line is not perpendicular to the field axis, but has a positive slope. Our refined phase diagram supports coexistence of phase I with superconductivity, as it has been also found in recent $\mu$SR experiments~\cite{khim2025}. 
\section{Results}
The main result of our work is the refined and most detailed $H-T$ phase diagram of the best-quality samples of \CRA\ to date for $H \parallel c$, which is shown in Fig.~\ref{fig1}. The points in the phase diagram were extracted from measurements of the $T$- and $H$-dependence of five thermodynamic quantities. In the following, we are going to show first measurements in zero field and then in magnetic field. We also explain how we extracted the points in Fig.~\ref{fig1} and comment on the nature of the phase transitions according to their signatures. The field $H$ was applied along the [001] direction in all measurements presented here.
\begin{figure}[t]
	\begin{center}
		\includegraphics[width=\columnwidth]{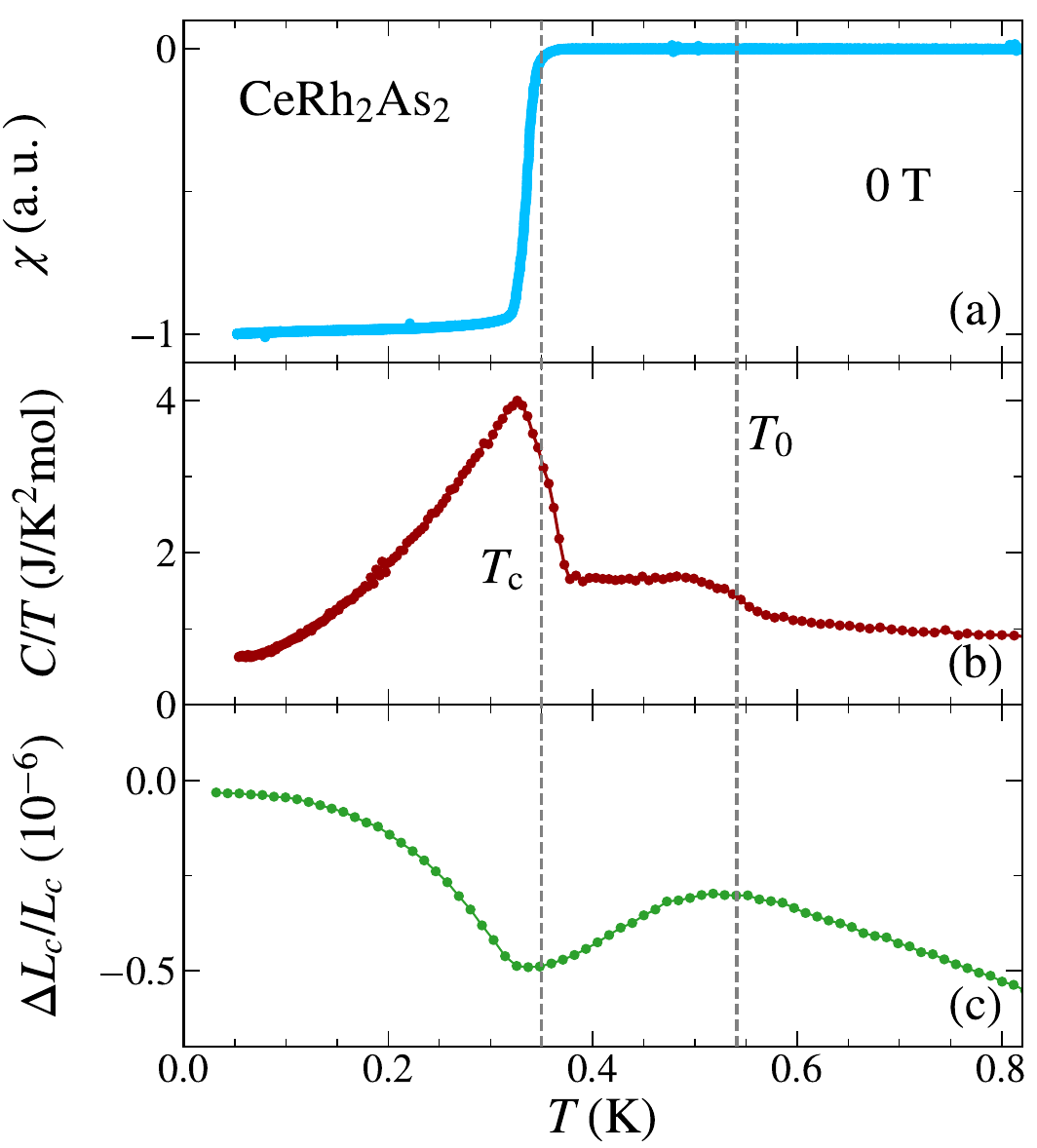}
		\caption{(a) $T$-dependence of the ac susceptibility $\chi(T)$. The gray dashed vertical lines indicate the position of the transition temperatures \Tc\ and \To\ estimated from the specific heat $C(T)/T$ plotted in (b). (c) $T$-dependence of the linear thermal expansion along the crystallographic $c$ axis $\Delta L(T)_{c}/L_{c}$.}
		\label{fig2}
	\end{center}
\end{figure}

The samples used in this study are from the highest-quality batch to date grown in Bi flux at the MPI-CPfS in Dresden~\cite{khim2021}. They were also used for the $\mu$SR study in Ref.~\cite{khim2025}. These samples have transition temperatures  \To\ = 0.54\,K and \Tc\ = 0.35\,K (cf. Figs.~\ref{fig1},\ref{fig2}). In Fig.~S1 of the Supplemental Material (SM) we compare the specific heat of three generations of samples with different quality. They are classified in three 'Groups' depending on the value of the critical temperatures and the sharpness of the transitions: The better the quality, the higher the critical temperatures and sharper the signature of the transitions, as expected in HF systems. Group 3 corresponds to the best quality. We also measured ac-susceptibility on a sample grown with Bi horizontal-flux method at the INTiBS in Wroc{\l}aw~\cite{chajewski2023,chajewski2024}. Although this particular sample has a slightly lower quality, comparable with samples of Group 2 used in Ref.~\cite{semeniuk2023}, it shows the very same features in field-dependent ac-susceptibility data (cf. Fig.~\ref{fig4}).  

In order to quantify the improvements in sample quality between Groups 2 and 3, we performed single-crystal X-ray diffraction on fragments extracted from Group 2 and 3 samples (see section SCXRD in the SM). Notably the Group 2 sample shows clear streaks of secondary reflections centred around the primary ones (Fig.~S2 of the SM). The probability distribution of the distance of the secondary reflections from the primary peaks along each principal axis (see Fig.~S3 of the SM) is 3 times broader in the Group 2 crystal compared to the Group 3 sample in all directions, with the disorder primarily occurring along the $c$ axis. Due to the layered nature of the structure, stacking faults can explain these features. However, our analysis can not exclude the presence of small ingrowths, with the lattice parameters similar to those of \CRA.

Figure~\ref{fig2} shows measurements of the ac-susceptibility $\chi(T)$, electronic specific heat coefficient $C(T)/T$ and linear thermal expansion along the crystallographic $c$ axis $\Delta L_{c}(T)/L_{c}$ in zero magnetic field. The high quality of the samples is indicated by the same transition temperatures measured in all quantities (cf. Refs.~\cite{khim2021} and~\cite{hafner2022}) and the sharp anomaly at the superconducting transition temperature \Tc. Both transition temperatures, \To\ = 0.54\,K and \Tc\ = 0.35\,K, were determined by equal entropy construction applied to the specific heat data, as it was done in previous works~\cite{khim2021,semeniuk2023,chajewski2024}.      

\begin{figure}[t]
	\begin{center}
		\includegraphics[width=\columnwidth]{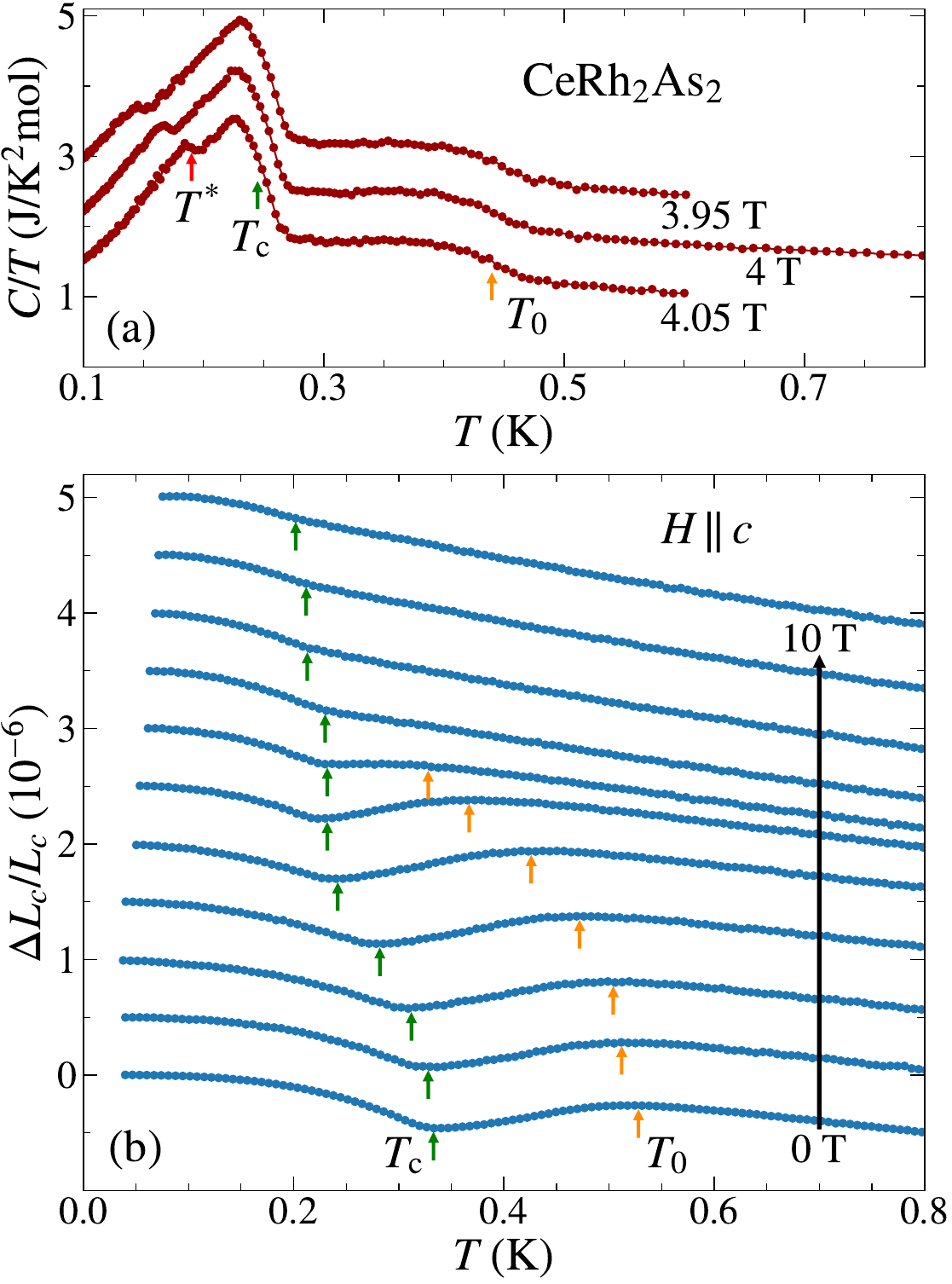}
		\caption{(a) Electronic specific heat coefficient $C(T)/T$ measured at 3.95, 4 and 4.05\,T, i.e., the fields near which the $T^{*}(H)$ line is observed (cf. Fig.~\ref{fig1}). The transition across this line from SC1 to SC2 is indicated by the red arrow at $T^{*}$. (b) Linear thermal expansion $\Delta L(T)_{c}/L_{c}$ measured at constant magnetic fields in steps of 1T from 0 to 10\,T. The curves are shifted upwards by $0.5\times 10^{-6}$ for clarity. The green and orange arrows indicate \Tc\ and \To, respectively.}
		\label{fig3}
	\end{center}
\end{figure}
%

Another interesting result is shown in Fig.~\ref{fig3}(a): The $T^{*}(H)$ line, which separates the SC1 from the SC2 phase, is almost perpendicular to the field axis (see Fig.~\ref{fig1}). The findings presented here show that the slope of $T^{*}(H)$ is clearly positive, which is consistent with the analysis of the jumps in the specific heat $\Delta C/T$ measured at the bicritical point (see SM of Ref.~\cite{khim2021}). In the $T$-dependent specific heat measurement at exactly 4.05\,T we indeed detect the phase transition from SC1+I to SC2+I at about 0.18\,K (red arrow in Fig.~\ref{fig3}(a)) indicating a small entropy change. The same feature rapidly shifts to lower temperatures when decreasing the field marginally from 4.05\,T to 3.95\,T. The transition points are plotted in Fig.~\ref{fig1} as green dots. The tiny $\mathrm{d}T^{*}/\mathrm{d}H$ slope observed in our experiments is predicted by theory~\cite{yoshida2012,khim2021}, but it could also be a signature of the recently proposed SC meron state~\cite{minamide2024}, which is a state in which the SC electron spins assume a skyrmion-lattice-like structure. 

Ref.~\cite{chajewski2024} observed additional features slightly below $T_{c} (H)$, sharp first-order peaks, which could be related to a vortex-lattice melting transition~\cite{schilling1997}. Despite the high quality of our samples, we could not see these sharp first-order phase transitions, likely because of the semi-adiabatic method used here~\cite{wilhelm2004}.

Both phase transitions at \Tc\ and \To\ and also their field dependence could be detected as kinks or maxima in $\Delta L(T)_{c}/L_{c}$ (marked by green and orange arrows in Fig.~\ref{fig3}(b), see also Fig.~S4 of the SM). These signatures indicate that both transitions remain second order also in magnetic field. The associated transition temperatures, extracted from the thermal expansion coefficient $\alpha_{c} = \frac{1}{L_{c}}\frac{d(\Delta L_{c})}{dT}$ are plotted in Fig.~\ref{fig1} as yellow triangles.
\begin{figure}[t]
	\begin{center}
		\includegraphics[width=\columnwidth]{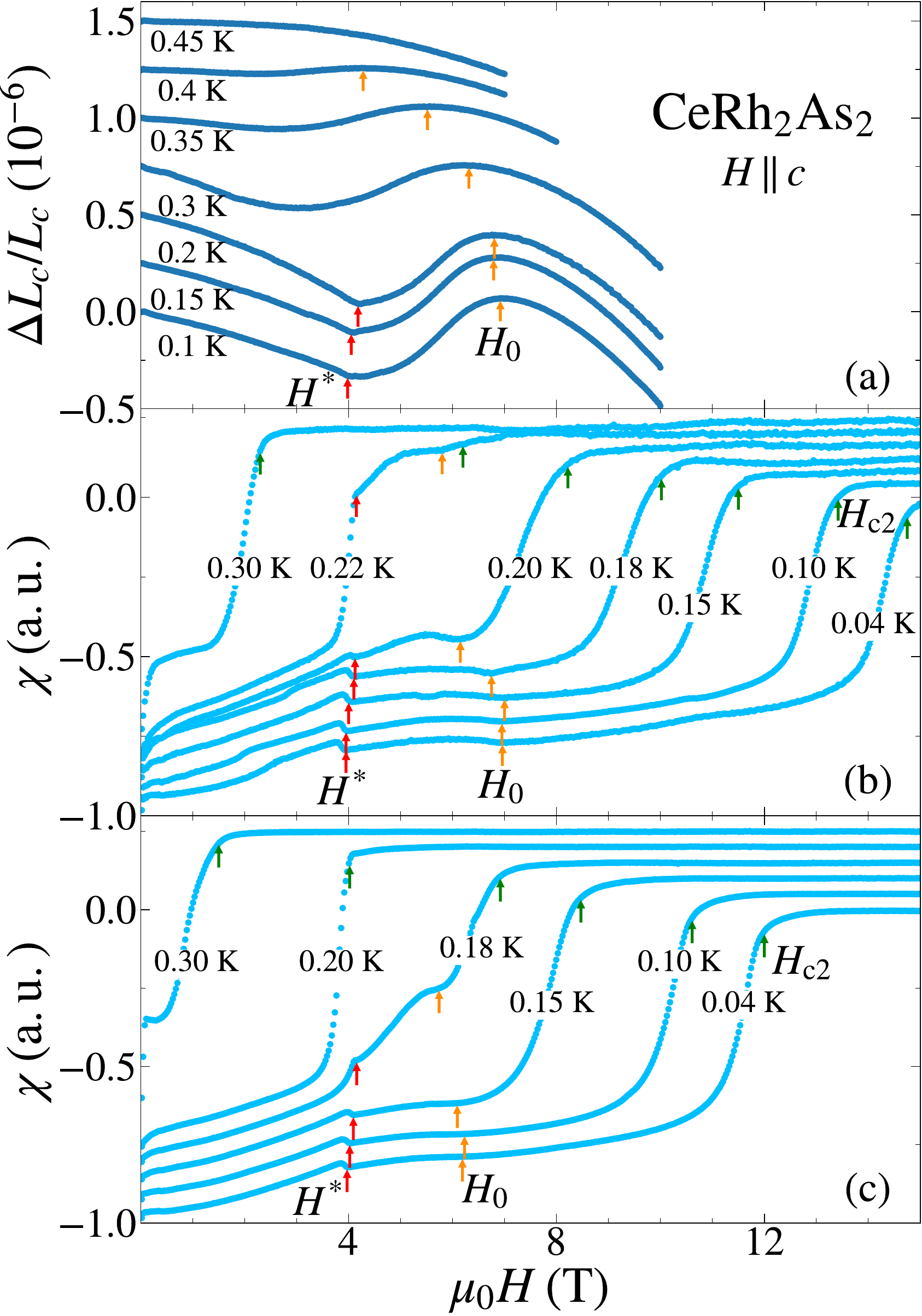}
		\caption{Field dependence of the (a) linear magnetostriction along the crystallographic $c$ axis $\Delta L_{c}(B)/L_{c}$ and of the (b),(c) ac susceptibility measured at different constant temperatures. Panel (b) shows measurements on the best quality sample (Group 3) whereas panel c) shows measurements on a slightly worse sample (Group 2) for comparison. We do plot here only sweep-up data since we could not see relevant hysteresis between sweeps up and down. The red and orange arrows indicate \Hstar\ and \Ho, respectively. The SC upper critical field $H_{c2}$ is indicated by green arrows. Curves in all panels are shifted for clarity.}
		\label{fig4}
	\end{center}
\end{figure}

\begin{figure}[t]
	\begin{center}
		\includegraphics[width=\columnwidth]{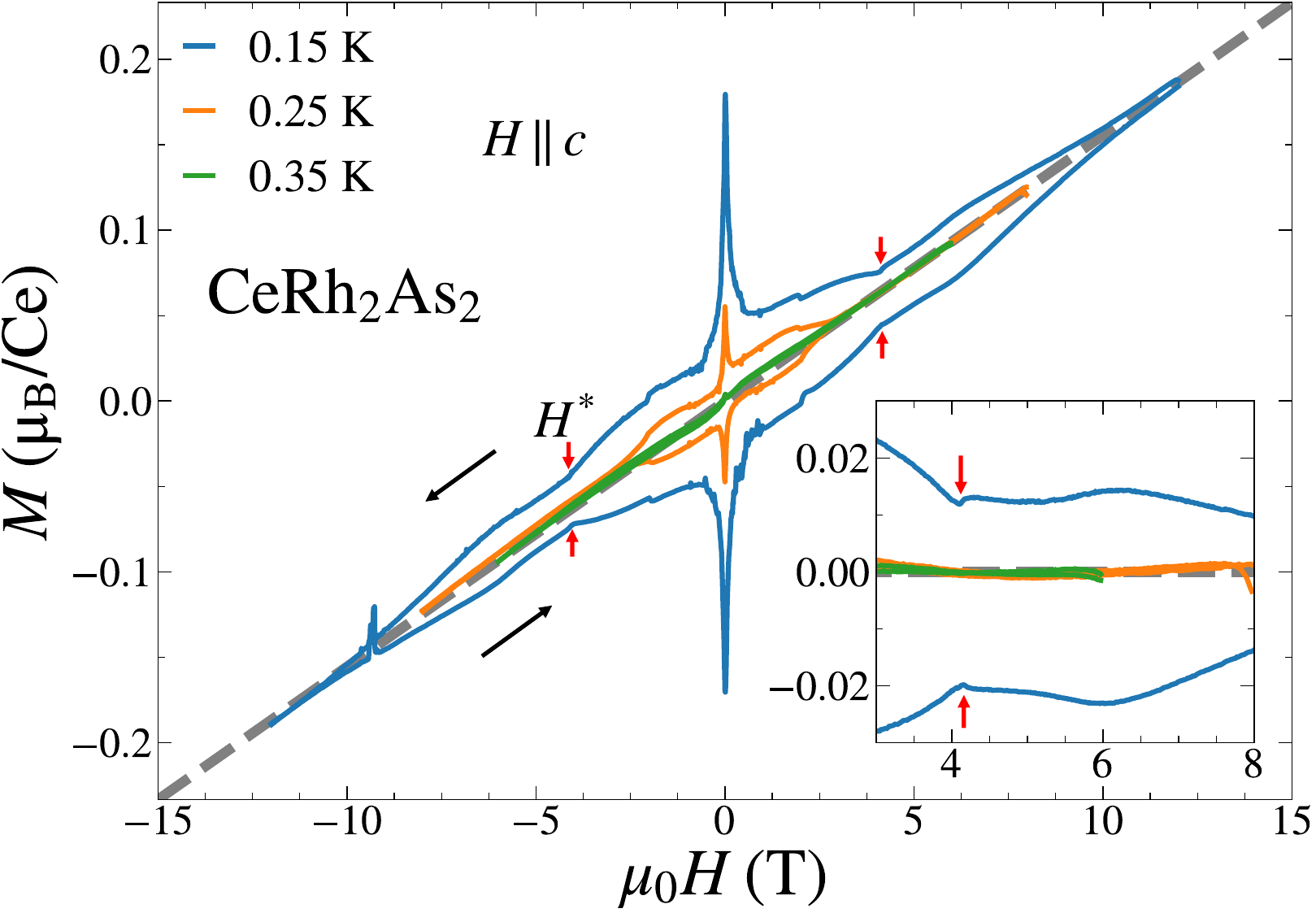}
		\caption{Field dependence of the magnetization $M(H)$ taken at three different temperatures. The hysteresis is due to vortex pinning and disappears above the SC transition temperature (cf. green curve). The red arrows mark the critical field $H^{*}$ while at $H_{0}$ we observe broad maxima. The step-like features at $\pm 2$\,T are artifacts due to the change in ramp rate of the magnet which influences the temperature controller.}
		\label{fig5}
	\end{center}
\end{figure}

The most relevant measurements are shown in Fig.~\ref{fig4}. These are $H$-dependent scans at constant temperatures of the ac-susceptibility $\chi(H)$ and linear magnetostriction $\Delta L_{c}(H)/L_{c}$. The associated transition fields, extracted from the magnetostriction coefficient $\lambda_{c} = \frac{1}{L_{c}}\frac{d(\Delta L_{c})}{dH}$ are plotted in Fig.~\ref{fig1} as orange squares. All measurements were carried out with up and down sweeps. Nevertheless, we could not observe any relevant hysteresis at any of the transition fields, i.e., only tiny hysteresis of the order of 10\,mT comparable with the remanent field of our superconducting magnets. A magnified view of the transitions at $H_{0}$ and $H^{*}$ in the susceptibility is shown in Fig.~S5 of the SM. This implies that all transitions are either second or weakly first order. We have marked all corresponding features in the data by arrows: Green arrows indicate the upper critical field of the SC2 phase, which we estimate to be when we recover 90\% of the normal state in the diamagnetic signal; The red arrows mark the transition fields $H^{*}$ between the SC1+I and the SC2+I phase. For $T \leq 0.22$\,K the signatures of the transition at $H^{*}$ are kinks in $\Delta L_{c}(H)/L_{c}$ and step-like features in $\chi(H)$, which correspond to the kinks observed in magnetization in Fig.~\ref{fig5} (red arrows). This behavior points to a second-order or weakly first-order phase transition. Moreover, $H^{*}$ effectively does not vary with sample quality (cf. Fig.~\ref{fig4}(b, c)). Robustness of the SC1-SC2 transition against disorder (see Fig.~S6 in SM) is expected for theories based on strong Rashba spin-orbit coupling if the ratio of Rashba and interlayer coupling does not change. Finally, the orange arrows indicate the critical field $H_{0}$ of phase I, which can be found also inside the phase SC2. The signatures here are minima in $\chi(H)$ (see Fig.~S5 of the SM) and maxima in $\Delta L_{c}(H)/L_{c}$. In this case, there is a difference in the values of $H_{0}$ between the data on the best-quality samples and those on the less pure sample, because $H_{0}$ is proportional to \To\ and therefore higher in the best-quality sample. It is worth noting that without $T$-dependent thermodynamic measurements and spectroscopic evidence for an ordered state in phase I~\cite{khim2025}, the signatures at $H_{0}(T)$ could be interpreted as a crossover rather than a phase transition. 

Another interesting observation is that in the non-SC state the ac susceptibility does not detect either a signature at $T_{0}$ or at $H_{0}$. The reason is probably that the AFM ordered moments are very small ($\leq 0.1\,$\muB) and mainly located within the basal plane~\cite{schmidt2024,ogata2024,juraszek2025}, whereas our measurements were carried out with ac field along the $c$ axis. At the same time, the likely reason why we see the signature within the superconducting state is an indirect one: Phase I presumably induces a small gap or a reconstruction of the Fermi surface that affects the screening currents of the superconducting state, which we see as an anomaly in the susceptibility.

The final phase diagram of \CRA\ for $H \parallel c$, shown in Fig.~\ref{fig1}, was constructed by using all features observed in our measurements and all information from other experiments. In Ref.~\cite{semeniuk2023}, we used a Ginzburg-Landau (GL) theory of coupled order parameters~\cite{imry1975,fernandes2010} to analyze the interplay between the SC order parameter $\Delta$ and the phase-I order parameter $\mathbf{Q}$ with a coupling term $\lambda\Delta^{2}\mathbf{Q}^{2}$, where $\lambda$ indicates the strength of the coupling as well as its nature, $\lambda > 0$ for competing and $\lambda < 0$ for supporting coupling. The model is well described in the Supplemental Material of Ref.~\cite{semeniuk2023}. We applied this theory to the phase boundaries intersecting at the tetracritical point 't' (see Fig.~\ref{fig1}). We were able to reproduce the observed phase boundaries for SC2 and I, and their slope change in the vicinity of 't'. The fits are shown as solid gray lines in Fig.~\ref{fig1}. In particular, with the addition of the newly discovered section of $T_{0}(H)$ line, the value of $\lambda$ is now shown to be between 20\% and 50\% of that necessary to induce a first order phase transition, i.e., the coupling is weakly competitive. As predicted in Ref.~\cite{semeniuk2023}, the $T_{0}(H)$ changes slope slightly when entering the SC2 phase and is almost vertical inside the SC2 phase. The weak coupling is corroborated by the $\mu$SR experiments~\cite{khim2025} which found evidence of microscopic coexistence between the SC phases and phase I.

\section{Conclusions}
We have presented a refined and complete version of the phase diagram of the HF superconductor \CRA\ for a magnetic field parallel to the tetragonal $c$ axis. In particular, we have found signatures of the phase I boundary line $T_{0}(H)$ within the superconducting phase SC2 confirming our presumption that there is a weak competing coupling between the order parameters of phase I and SC2. Our results corroborate recent $\mu$SR results which suggest microscopic coexistence of phase I with superconductivity. Moreover, we find evidence of the putative first-order $T^{*}(H)$ transition line in $T$-dependent specific heat measurements. We could not detect relevant hysteresis in all $H$-dependent measurements across this line. This suggests that $T^{*}(H)$ transition is either second or weakly first order.

\begin{acknowledgments}
We are indebted to T. Lorenz, M. Garst for useful discussions. This work is also supported by the joint Agence National de la Recherche and DFG program Fermi-NESt through Grants No. GE602/4-1 (C. G. and E. H.). Additionally, E. H. acknowledges funding by the DFG through CRC1143 (project number 247310070) and the W\"urzburg-Dresden Cluster of Excellence on Complexity and Topology in Quantum Matter — ct.qmat (EXC 2147, project ID 390858490). This work is supported by the ERC grant (Ixtreme, GA 101125759). S. K. was supported by the Max Planck Society and the Deutsche Forschungsgemeinschaft (DFG, German Research Foundation) - KH 387/1-1. A.R. and M.P were supported by the Engineering and Physical Sciences Research Council (grants EP/P024564/1, EP/V049410/1 and EP/L015110/1). M. P. acknowledges financial support by the International Max Planck Research School for Chemistry and Physics of Quantum Materials (IMPRS-CPQM).
\end{acknowledgments}
\bibliography{khanenko_prbl_2025.bib}
\bibliographystyle{apsrev4-2}
\end{document}